\documentclass[12pt,amsfonts]{article} % New Coherent States based on Affine Quantum Fields  Sept. 7, 2009
\usepackage{amsfonts}
\def\one{1\hskip-.37em 1}

\def\half{\textstyle{\frac{1}{2}}}

\def\H{{\cal H}}

\def\p{\phi}
\def\th{\theta}
\def\H{{\cal H}}

\def\t{\textstyle}

\def\ra{\rightarrow}
\def\tint{{\textstyle\int}}

\def\s{\hskip.08em}

\def\d{\partial}
\def\o{\overline}

\def\b{\begin{eqnarray*}}  %takes no eqn numbers
\def\e{\end{eqnarray*}}    %takes no eqn numbers
\def\bn{\begin{eqnarray}}  %takes eqn numbers
\def\<{\langle}
\def\>{\rangle}

\def\no{\nonumber}

\def\k{\kappa}
\def\{{\lbrace}
\def\}{\rbrace}
\title{New Affine Coherent States based on \\Elements of Nonrenormalizable\\
Scalar Field Models}
\author{John R. Klauder\footnote{klauder@phys.ufl.edu}
\\Department of Physics and \\Department of Mathematics\\
University of Florida\\
Gainesville, FL 32611-8440}
\date{ }
\begin{document}
\maketitle
\begin{abstract}
Recent proposals for a nontrivial quantization of covariant, nonrenormalizable,
self-interacting, scalar quantum fields have emphasized the importance of quantum
fields that obey affine commutation relations rather than canonical commutation
relations. When formulated on a spacetime lattice, such models have a lattice version of the
associated ground state, and this vector is used as the fiducial vector for the
definition of the associated affine coherent states, thus ensuring that in the
continuum limit, the affine field operators are compatible with the system
Hamiltonian. In this article, we define and analyze the associated affine
coherent states as well as briefly
review the author's approach to nontrivial formulations of such
nonrenormalizable models.

\end{abstract}
\section{Introduction and Overview}
The author has long been interested in finding a meaningful and nontrivial quantization
of nonrenormalizable quantum field theories \cite{kla17}. Quantum gravity is the most
important example of this
kind, but it is kinematically simpler to try to understand nonrenormalizable, self-interacting scalar
fields as test cases initially. The subject of this article continues that quest by
proposing coherent states based on affine field algebras associated with a recent scheme
to quantize scalar nonrenormalizable models. Suitable coherent states have the virtue of bridging
the classical-quantum divide, and thus offer a useful tool in the study of such problems.

In Sec.~2 we introduce the concept of affine coherent states first starting with simple
few degree-of-freedom systems. Next, we extend that discussion to affine fields
in a 1-dimensional and later to an $s$-dimensional
Euclidean space. To get a better handle on such representations, we study cutoff
field examples by replacing space by a finite, periodic lattice of points
with $L<\infty$ points on a side each separated by a lattice spacing $a>0$. Each point on the
lattice carries a set of affine fields that obey a discrete form of the affine field algebra
and thus represent a finite but large number of independent, one-dimensional affine systems.
In the continuum limit, of course, such systems become affine field algebras, but studying them
before taking the continuum limit offers certain advantages. In addition, we choose very
special vectors to serve as fiducial vectors for the associated affine coherent states. These
fiducial vectors are chosen as ground states for suitable model field theories that
in Sec.~3 are also formulated on spacetime lattices with similar properties. These quantum field
models have been chosen  to achieve a nontrivial quantization of quartic, self-interacting,
covariant, nonrenormalizable models, specifically the so-called $\varphi^4_n$ models in
spacetime dimensions $n\ge5$. These models are sketched in Sec.~3 to an extent necessary so as to
understand the choice of vectors selected  as fiducial vectors for the discussion in Sec.~2.
A brief commentary appears in Sec.~4 setting the present work in the context of efforts to
study quantum gravity.

\section{New Affine Coherent States}
\subsection*{Single degree of freedom}
For a single degree of freedom, the canonical, self adjoint, quantum variables are $P$ and $Q$, which obey the canonical
commutation relation $[Q,P]=i\hbar\one$. The usual irreducible representation of these operators involves spectra for both operators that runs from $-\infty$ to $+\infty$. The affine variables are formally obtained
from the canonical ones by multiplying the canonical commutation relation by the operator $Q$, which leads to
the relation
  \b Q\s[Q,P]=i\hbar\s Q=[Q,(QP+PQ)/2]\equiv [Q,D]\;, \e
an expression that introduces the affine quantum variables $Q$ and $D$ which satisfy the affine commutation
relation $[Q,D]=i\hbar\s Q$. It is noteworthy that the affine commutation relation has three inequivalent, irreducible representations with self-adjoint affine variables: one for which $Q>0$, one for which $Q<0$, and one for which $Q=0$. For the first two representations, $D$ has a spectrum that covers the whole real line. For the irreducible representation for which $Q>0$, a suitable set of coherent states may be given by
   \b |p,q +\>\equiv e^{\t ip\s Q/\hbar}\,e^{\t -i\ln(\s q\s)\,D/\hbar}\,|\eta+\>\;, \e
    where $p\in\mathbb{R}=(-\infty,\infty)$, $q\in\mathbb{R}^+=(0,\infty)$, and the unit vector $|\eta+\>$ is called the fiducial vector.
    In this formulation $q$ is dimensionless while $p$ has the dimensions of $\hbar$. In the representation in which $Q$ is diagonalized, i.e.,
    $Q\s|x\>=x\s|x\>$, $x>0$, , it follows that the function $\eta_+(x)=\<x|\eta+\>$ is supported only on the positive real line. One possible choice of fiducial vector is given by
       \b \eta_+(x) =N_+\,\exp[\s-(B/x^2)-C(x-1)^2/\hbar\s]\;,\hskip1cm x>0\;, \e
       where $B>0$, $C>0$, and $N_+$ normalizes the expression so that $\|\s|\eta+\>\s\|=1$.  If instead we were interested in
       the second irreducible representation for which $Q<0$, then a suitable set of coherent states
       would be given (for $q<0$) by
   \b |p,q -\>\equiv e^{\t ip\s Q/\hbar}\,e^{\t -i\ln(\s -q\s)\,D/\hbar}\,|\eta-\>\;, \e
   In this case $Q\s|x\>=x\s|x\>$, $x<0$, and we may choose
      \b \eta_-(x)=N_-\,\exp[\s-(B/x^2)-C(x+1)^2/\hbar\s]\;, \e
      from which we conclude that $\|\s|\eta-\>\s\|=1$.
      \subsubsection*{Reducible representation}

      Finally, when we wish to include both irreducible representations, where $Q>0$ and $Q<0$, we may choose
      a set of coherent states---now for $p\in\mathbb{R},\,q\in\mathbb{R}\setminus0$---given by\texttt{\texttt{}}
      \b |p,q\pm\>\equiv \s\theta(q)\,|p,q+\>+\s\theta(-q)\,|p,q-\>\;,\e
       where $\theta(q)\equiv1$ if $q>0$ and $\theta(q)\equiv0$ if $q<0$. It follows that
        \b \|\s|p,q\pm\>\s\|^2=\theta(q)\s\|\s|p,q+\>\s\|^2+\theta(-q)\s\|\s|p,q-\>\s\|^2=1\;, \e
        and thus the coherent states $|p,q\pm\>$ are all normalized vectors.

      It is  useful to review some algebraic properties of affine variables and some properties of affine coherent state matrix elements. In particular, if we introduce ($q\not=0$)
          \b A[p,q]\equiv e^{\t ip\s Q/\hbar}\,e^{\t -i\ln(\s |q|\s)\,D/\hbar}\;,  \e
       we have the relations
         \b    A[p,q]^\dagger\s D\s A[p,q]\hskip-1.3em&&= D+ p\s |q|\s Q \;,\\
         A[p,q]^\dagger\s P\s A[p,q]\hskip-1.3em&&= |q|^{-1} P+p\;,\\
         A[p,q]^\dagger\s Q\s A[p,q]\hskip-1.3em&&=  |q|\s Q\;,\e
         and
         \b
                &&\<p,q\pm|\s B(D,P,Q)\s|p,q\pm\>=\th(q)\<\eta+|\s B(D+p\s q\s Q,q^{-1} P+p,q\s\s Q)\s|\eta+\>\\
                &&\hskip1cm +\th(-q)\<\eta-|\s B(D+p\s (-q)\s Q,(-q)^{-1} P+p,(-q)\s\s Q)\s|\eta-\>\;.\e
               Introducing $\<(\s\cdot\s)\>_\pm\equiv \<\eta\pm|(\s\cdot\s)\s|\eta\pm\>$, it follows that
                \b\<p,q\pm|\s Q\s |p,q\pm\>=\th(q)\s q\s\<Q\>_++\th(-q)\s(-q)\s\<Q\>_-\;, \e
                 and we note that if we choose the parameters $B$ and $C$ such that $\<Q\>_+=1$, it follows that $\<Q\>_-=-1$, and we find that
                  \b \<p,q\pm|\s Q\s|p,q\pm\> = \th(q)\s q-\th(-q)\s (-q)\equiv q\;, \e
                  a desirable value indeed. Additionally, for whatever values we assign to $B$ and $C$ it follows
                  that
                     \b \<p,q\pm|\s Q\s|p,q\pm\>= q+O(\hbar)\;, \e
                  and, furthermore, it is not difficult to show that
                    \b \<p,q\pm|\s Q^r\s|p,q\pm\>=q^r+O(\hbar)\;,  \e
                    for all $r$.

                   In addition,
                   \b  \<p,q\pm|\s D\s|p,q\pm\>\hskip-1.3em&&=\th(q)[\s p\s q\s\<Q\>_++\<D\>_+]+\th(-q)[\s p\s(-q)\<Q\>_-+\<D\>_-]  \\
                                  &&=p\s q \;,\e
                   where we have chosen the physically natural values that  $\<D\>_\pm=0$, which follow
                   directly when the parameters $B$ and $C$ are both real. These expectation values
                   enable us to identify $q$ as the mean value of $Q$  and $p\s q$ as the mean value of
                   $D$ in the coherent states. This identification introduces a natural connection
                   between the quantum and classical variables, but in no sense is
                   this particular connection required.

                    \subsection*{Alternative affine coherent states}
                    The construction of $|p,q\pm\>$ entails two distinct parts proportional to $\theta(q)$ and
                    $\theta(-q)$, respectively. If $J$ degrees of freedom are defined this way, then $2^J$
                    disjoint parts are involved, which becomes very large when $J\gg1$. To avoid this aspect, we next focus on an alternative
                    construction.

                   In this new version, we choose to give up some features of the set
                   of coherent states defined above and instead define
                     \b |p,q\>\equiv e^{\t ip\s Q/\hbar}\,e^{\t -i\ln(\s|q|\s)\,D/\hbar}\,|\eta\>\;,\e
                     where the spectrum of $Q$ is taken to be the whole real line.
                      Observe, in this case, that the state $|p,-q\>=|p,q\>$, namely the coherent states
                     for  $q>0$ and $q<0$ are identical to each other. Additionally, in this
                     case we consider $Q\s|x\>=x\s|x\>$. where $-\infty<x<0$ and $0<x<\infty$, and
                     $\eta(x)=\<x|\eta\>$ is conveniently  chosen as an even function, i.e., $\eta(-x)=\eta(x)$.
                     Consequently, using the abbreviation
                     $\<(\cdot)\>\equiv \<\eta|(\cdot)|\eta\>$, we find that
                       \b \<p,q|\s Q\s|p,q\>=|q|\s\<\s  Q\s\>=0 \e
                       for all $(p,q)$, and likewise $\<p,q|\s Q^{2r+1}\s|p,q\>=0$ for any odd power.

                     However, let us instead consider
                     \b \<p,q|\s Q^2\s|p,q\>=q^2\,\<\s Q^2\s\>\;, \e
                     which, along with the modest requirement that $\<\s Q^2\s\>=1$, leads to
                     \b \<p,q|\s Q^2\s|p,q\>=q^2 \e
                     from which we can conclude that $q^2$ is the mean value of $Q^2$.
                     By a suitable choice of the fiducial vector $|\eta\>$, we can also arrange that
                       \b \<p,q|\s Q^{2r}\s|p,q\>=q^{2r}+O(\hbar)\;.  \e

                 In the same states $|p,q\>$, it follows that
                 \b \<p,q|\s B(D,P,Q)\s|p,q\>=\<\s B(D+p\s|q|\s Q,|q|^{-1}P+p,|q|\s Q)\s\>\;.\e
If the function $\eta(x)=\<x|\eta\>$ is real, then besides $\<\s Q\s\>=0$ it also follows that
$\<\s D\s\>=0$ and $\<\s P\s\>=0$, which leads to
      \b \<p,q|\s D\s|p,q\>=0\;,\hskip1cm \<p,q|\s P\s|p,q\>=p\;,\e
      and thus implies that $p$ is the mean value of $P$ in the coherent state $|p,q\>$.
      Additionally, we see that
        \b \<p,q|\s P^2\s|p,q\>=\<\s (|q|^{-1}\s P+p)^2\s\>=p^2+q^{-2}\s\<\s P^2\s\>\;,\e
        where the factor $\<P^2\>=\hbar^2\s c$ for a dimensionless constant $c>0$.

        With the foregoing expectation values, we can argue that
        the expression given by
        \b \<p,q|\s \H\s|p,q\>\hskip-1.2em&&\equiv\<p,q|\s\s \half\s (P^2+\omega^2\s Q^2)+ \lambda \s Q^4\s\s|p,q\>\\
          &&=\half\s(p^2+\hbar^2\s c/q^2+\omega^2\s q^2)+\lambda_0 \s q^4\\
          &&=\half(p^2+\omega^2\s q^2)+\lambda\s q^4+O(\hbar) \e
          seems to provide a reasonable connection between suitable quantum and
          classical Hamiltonians.

        Traditionally, when dealing with coherent states, one also speaks about a resolution of unity in the form
       \b  \one=\int |p,q\>\<p,q| \,w(p,q)\,dp\s dq\;, \e
       for some weight function $w(p,q)$ that is positive almost everywhere. However, in this article we shall not focus on this form of the resolution of unity. Instead, we shall rely on the fact that the coherent state
       overlap function $\<p'',q''|p',q'\>$ serves as a reproducing kernel for a reproducing kernel Hilbert
       space representation of the underlying abstract Hilbert space $\frak{H}$; see \cite{aron}.

\subsection*{Many dimensional affine coherent states}
We now extend the preceding analysis to a discussion of many dimensional affine coherent states. Consider
a set of $J$ ($J$ being {\it odd}) independent affine fields such as $Q_j$ and $D_j$, $j=0,\pm1,\pm2,\ldots,\pm                          J^*$, where $J^*\equiv(J-1)/2$, and the only nonvanishing commutator is given by
 \b [Q_l,D_j]=i\hbar\s\delta_{l,j}\s Q_l\;, \e
  and for each $j$ the spectrum of $Q_j$ is the entire real line save for zero. The coherent states for this system are taken as
    \b |p,q\>\equiv e^{\t i\s \Sigma_j p_j\s Q_j/\hbar}\,e^{\t -i\Sigma_j\s \ln(|q_j|)\s D_j/\hbar}\,|\eta\>\;. \e
    In terms of the states $|x\>$, where $Q_j\s|x\>=x_j\s|x\>$, for all $j$, it follows that
    \b \<x|p,q\>=[\s\Pi_j|q_j|^{-1/2}]\,e^{\t i\Sigma\s p_j\s x_j/\hbar}\,\eta(x/|q|)\;, \e
    where $\eta(x/|q|)\equiv\eta(x_{-J^*}/|q_{-J^*}|,\dots,x_0/|q_0|,\ldots,x_{J^*}/|q_{J^*}|)$.

    As a first example, we assume (with $\hbar=1$) that
    \b \eta(x)=\<x|\eta\>=N\,{e^{\t -\half\s\omega \Sigma_j\s x^2_j}}  \;. \e
    In this case,
    \b \<x|p,q\>=N \,[\s\Pi_j|q_j|^{-1}]^{1/2}\,{e^{\t i\Sigma_j p_jx_j -\half\omega \Sigma_j\s (x^2_j/q_j^2)}} \;, \e
    and the overlap of two such coherent states is given by
    \b \<p',q'|p,q\>={\t\prod}_{j=-J^*}^{J^*}\frac{[2\s|q'_j|^{-1}\s|q_j|^{-1}\s]^{1/2}}{[\s q'^{-2}_j+q_j^{-2}\s]^{1/2}}
    \,e^{\t -(1/2\omega)(p'_j-p_j)^2/[\s q'^{-2}_j+q_j^{-2}\s]}\;.\e

    Let us extend our present example to an infinite number of degrees of freedom, for which $J\ra\infty$.
    This leads to the expression
      \b \<p',q'|p,q\>={\t\prod}_{j=-\infty}^\infty\frac{[2\s|q'_j|^{-1}\s|q_j|^{-1}\s]^{1/2}}{[\s q'^{-2}_j+q_j^{-2}\s]^{1/2}}
    \,e^{\t -(1/2\omega)(p'_j-p_j)^2/[\s q'^{-2}_j+q_j^{-2}\s]}\;,\e
    which provides a well-defined product representation for affine coherent states
    provided that the variables $\{p'_j,q'_j\}_{j=-\infty}^\infty$ and $\{p_j,q_j\}_{j=-\infty}^\infty$
    are well chosen.

    \subsubsection*{Representation for a field}
    A one-dimensional affine field theory involves field operators that satisfy
    the affine commutation relation
      \b [\s\varphi(x),\rho(y)\s]=i\s\hbar\s\delta(x-y)\s\varphi(x)\;, \hskip1cm x,y\in\mathbb{R}\;,\e
      where $\varphi(x)$ is the generalization of $Q$ and $\rho(y)$ generalizes $D$. Let us regularize this field
      formulation by introducing  a one-dimensional lattice space with $J<\infty$ lattice points (again with $J$
      conveniently chosen as an {\it odd} number) each separated by a
      lattice spacing $a>0$, which leads to a regularized affine field representation given by fields
      $\varphi_k$ and $\rho_k$, where $x$ has been replaced by the integer $k$ and $x=k\s a$. Here,
      $-J^*\le k\le J^*$, and $J^*\equiv(J-1)/2$ as before. These operators
      obey the affine commutation relation given in the form
          \b [\s\varphi_j,\rho_k\s]=i\s\hbar\s a^{-1}\s\delta_{j,k}\s\varphi_j\;.\e
          The coherent states for such a regularized field are given by
             \b |p,q\>\equiv e^{\t i\Sigma_j p_j\varphi_j\s a/\hbar}\,e^{\t-\Sigma_k\s\ln(|q_k|)\s \rho_k\s a/\hbar}\,|\eta\>\;.\e

             For our {\bf first example}, we choose $|\eta\>$ so that (with $\hbar=1$)
                  \b \<\phi|\eta\>=M\,e^{\t -\half\s\omega\s\Sigma_k\s\phi_k^2\s a}\;; \e
                  here, the vector $|\p\>$ replaces $|x\>$ as used before, where $\varphi(x)\s|\p\>=\p(x)\s|\p\>$. In the present
                case the overlap function of two coherent states becomes
                  \b  \<p',q'|p,q\>={\t\prod}_{j=-J^*}^{J^*}\frac{[2\s|q'_j|^{-1}\s|q_j|^{-1}\s]^{1/2}}{[\s q'^{-2}_j+q_j^{-2}\s]^{1/2}}
    \,e^{\t -(1/2\omega)(p'_j-p_j)^2\s a/[\s q'^{-2}_j+q_j^{-2}\s]}\;.\e

    Next we investigate a possible continuum limit in which $J^*\ra\infty$, $a\ra0$, and initially we require that
    $(2\s J^*+1)\s a=J\s a\equiv X$ may be large but finite; a subsequent limit in which $X\ra\infty$ is taken later.
    Moreover, in this limit we also insist that $p_j\ra p(x)$and $q_j\ra q(x)$,  and that both functions are {\it continuous}. To ensure that we focus on the representations induced by the given fiducial vector,
    we restrict attention to functions $p(x)$ that have compact support and functions $q(x)$ such that
    $\ln|q(x)|$ also has compact support, or stated otherwise, $|q(x)|=1$ outside a compact region. It is clear that the exponential factor exhibits a
    satisfactory continuum limit given by
         \b  \Sigma_j\s(p'_j-p_j)^2\s a/[\s q'^{-2}_j+q_j^{-2}\s]\ra \tint\s[p'(x)-p(x)]^2/
         [q'(x)^{-2}+q(x)^{-2}]\,dx\;.\e
         However, the continuum limit of the prefactor turns out to be
         {\it identically zero} unless $|q'(x)|=|q(x)|$ for all $x$! This implies that $\<p',q'|p,q\>=0$ whenever
         $|q'(x)|\not\equiv |q(x)|$ (as befits a {\it non$\s$}separable Hilbert space!), and thus the operator
         representation with this fiducial vector has turned out to be highly singular. Stated otherwise, the affine field algebra fails to have an acceptable continuum limit for a strictly Gaussian fiducial vector with the indicated form.

    As a {\bf second example}, suppose that
 \b \eta(\p)=\<\p|\eta\>=N\frac{e^{\t -\half\s\omega \Sigma_j\s \p^2_j\s a}}{[\s\Pi_j\s|\p_j|\s]^{(J-1)/2J}}\;. \e
    This unusual expression leads to a normalizable fiducial vector, i.e.,
        \b  N^2 \int  \frac{e^{\t -\omega \Sigma_j\s \p^2_j\s a}}{[\s\Pi_j\s|\p_j|\s]^{(J-1)/J}}\,\Pi_j\s d\p_j=1\;,\e
        the finiteness of which is clear.

       For the second example, we find that
        \b \<\p|p,q\>=N\s[\s\Pi_j\s|q_j|^{-1}\s]^{1/2J}\,  \frac{e^{\t i\Sigma_k p_k\s \p_k\s a -\half\omega \Sigma_j\s (\p^2_j/q_j^2)\s a}}     {[\s\Pi_j\s|\p_j|\s]^{(J-1)/2J}}\;, \e
 and the coherent state overlap function $\<p',q'|p,q\>$ for the second example is given by
        the integral
           \b \<p',q'|p,q\>\hskip-1.4em&&=N^2\s[\Pi_k|q'_k|^{-1}\s|q_k |^{-1}]^{1/2J}\,\int e^{\t -i\Sigma_k(p'_k-p_k)\p_k\s a}\\ &&\hskip1em \times e^{\t -
           \half\s\omega\s\Sigma_k\s \p_k^2\s[\s q'^{-2}_k+q^{-2}_k]\s a}\,\frac{1}{[\s\Pi_j\s|\p_j|\s]^{(J-1)/J}}
           \,\Pi_kd\p_k\\
           &&= N^2\s{\t\prod_k}\frac{[2\s|q'_k|^{-1}\s|q_k |^{-1}]^{1/2J}}  {[\s q'^{-2}_k+q^{-2}_k\s]^{1/2J}}\\ &&\hskip1em\times\int\frac{e^{\t -i\Sigma_k(p'_k-p_k)\p_k\s a/[(\s q'^{-2}_k+q^{-2}_k\s)/2]^{1/2}}
           e^{\t-\omega\s\Sigma_k\s \p^2_k\s a}}   {[\s\Pi_j\s|\p_j|\s]^{(J-1)/J}}\,\Pi_kd\p_k\;. \e
           The normalization factor in this relation follows from the fact that
           \b 1=\<p,q|p,q\>=N^2\s\int\frac{
           e^{\t-\omega\s\Sigma_k\s \p^2_k\s a}}   {[\s\Pi_j\s|\p_j|\s]^{(J-1)/J}}\,\Pi_kd\p_k\;.\e

      In considering the continuum limit, we again focus on
      continuous functions $p(x)$ and $\ln|q(x)|$ that have compact support, and we initially restrict attention to a finite overall interval $X=J\s a$. In this case, the new prefactor involves the $1/J^{\rm th}$ root of the previous prefactor, and this new version leads to an acceptable continuum limit. In particular,
         it is possible to evaluate the prefactor itself exactly---which also equals the coherent state matrix
        element $\<p,q'|p,q\>$---as
           \b \<p,q'|p,q\>\hskip-1.5em&&= \lim_{a\ra0}{\t\prod_j}\frac{[2\s|q'_j|^{-1}\s|q_j |^{-1}]^{1/2J}}  {[\s q'^{-2}_j+q^{-2}_j\s]^{1/2J}}\\
           &&\hskip-2em=\lim_{a\ra0}\exp[-1/(2J\s a)\Sigma_j\{\ln|q'_j|+\ln|q_j|+\ln[(q_j^{'-2}+q_j^{-2})/2]\}\s a\s]\\
           && \hskip-2em= \exp[-(1/2X)\tint \{\s\ln|q'(x)|+\ln|q(x)|+\ln[(q'(x)^{-2}+q(x)^{-2})/2]\s\}\,dx]\;. \e
     For arguments of compact support, it follows in the limit that $X\ra\infty$,  that the prefactor
     becomes {\it unity} and thus $\<p,q'|p,q\>=1$ for all arguments! While unusual, this result is
     perfectly acceptable from a reproducing kernel point of view. (Indeed, this result is surely more acceptable
     than the conclusion for the first example where $\<p,q'|p,q\>$ was identically zero except when
     $|q'(x)|=|q(x)|$ for all $x$.) Moreover, for the second example, the expression $\<p',q'|p,q\>$ is well defined and is  continuous in its labels in the continuum limit when $X$ is finite as well as in the
     further limit that $X\ra\infty$. Thus, although we have not explicitly evaluated the coherent state
     overlap function in the general case for the second example, it is clearly a continuous function of positive type suitable to be a reproducing kernel for the (separable) Hilbert space of interest.

  \subsection*{Construction of field theoretic affine coherent states}
  In light of the foregoing discussion, it is but a small step to introduce the affine coherent
  states of interest in the study of nonrenormalizable scalar field models. In the rest of this section,
  we introduce the ground state of our proposed models and define the associated affine coherent states
  using that ground state as the fiducial vector; in the following section we give a brief discussion that
  outlines the motivation for the particular choice of the ground state for these models.

      Our analysis is aimed at a field theory model regularized by a spacetime lattice that has one
time dimension and $s$ space dimensions. Focusing on the spatial aspects, we put $L<\infty$ lattice points in each of the $s$ directions
with a lattice spacing of $a>0$. Each lattice point is labeled by a multi-integer $k=(k_1,k_2,\ldots,k_s)$,
where each $k_j\in\mathbb{Z}=\{0,\pm1,\pm2,\cdots\}$. The total number of lattice points in a spatial slice of the lattice is given by $N'\equiv L^s$, a factor which plays the role that $J$ played in examples one and two above. In the representation in which the field operator
is diagonal, there is a field associated with each lattice point, $\p_k$, much as in examples one and two above,
save for the fact that now the label $k$ is $s$-dimensional. In the present case, the ground state of the system
Hamiltonian is chosen (with $\hbar=1$) as
   \b \Psi_0(\p)= M \frac{e^{\t -\half \Sigma'_{k,l}\p_k\s A_{k-l}\s \p_l\s a^{2s}-\half\s W(\p\s a^{(s-1)/2})}}{\Pi'_k[\s\Sigma'_l\s
   J_{k,l}\s\p^2_l\s]^{(N'-1)/4N'}}\;, \e
where primes on products and sums signify they apply only to a given spatial slice.
In this expression, $A_{k-l}$ is a numerical matrix proportional to $a^{-(1+s)}$ and $J_{k,l}=1/(2s+1)$ for
$l=k$ and when $l$ is any of the $2s$ nearest neighbors of $k$ in the same spatial slice; otherwise $J_{k,l}=0$.
The role of $J_{k,l}$ is to provide a local average of field values in the sense that ${\o{\p^2_k}}\equiv
\Sigma_l\s J_{k,l}\s\p^2_l$, which means that the denominator exhibits an integrable singularity for the
ground state distribution even in the limit $N'\ra\infty$. The $A$ factor in the exponent and the $J$ factor in the denominator well
represent the functional form of the ground state for large and small field values,
respectively; the unspecified
function $W$ (discussed in the following section) is needed to modify intermediate field values.

The affine coherent states for this fiducial vector are given by
  \b \<\p|p,q\>\hskip-1.5em&&=M\s\frac{ \Pi'_k[\s|q_k|^{-1/2}\s]}{\Pi'_k[\s\Sigma'_l\s
   J_{k,l}\s(\p^2_l/q^2_l)\s]^{(N'-1)/4N'}}\\
  &&\hskip-2em\times {e^{\t i\Sigma'_k p_k \p_k\s a^s-\half\Sigma'_{k,l}(\p_k/|q_k|)\s A_{k-l}\s
  (\p_l/|q_l|)\s  a^{2s}-\half\s W((\p/|q|)\s a^{(s-1)/2})}}\;, \e
which leads to the overlap expression
   \b \<p',q'|p,q\>\hskip-1.5em&&=M^2\,\Pi'_k\{[\s|q'_k|^{-1/2}\s][\s|q_k|^{-1/2}\s]\}\,
   \int \Pi'_k\s d\p_k\,e^{\t-i\Sigma'_k(p'_k-p_k)\p_k\s a}\\
      &&\times{e^{\t -\half\Sigma'_{k,l}(\p_k/|q'_k|)\s A_{k-l}\s (\p_l/|q'_l|)\s a^{2s}
      -\half\s W((\p/|q'|)\s a^{(s-1)/2})}}\\
      &&\times {e^{\t -\half\Sigma'_{k,l}(\p_k/|q_k|)\s A_{k-l}\s (\p_l/|q_l|)\s
      a^{2s}-\half\s W((\p/|q|)\s a^{(s-1)/2})}}\\
      &&\times \frac{1}{{\Pi'_k[\s\Sigma'_l\s
   J_{k,l}\s(\p^2_l/q'^2_l)\s]^{(N'-1)/4N'}}\;{\Pi'_k[\s\Sigma'_l\s
   J_{k,l}\s(\p^2_l/q^2_l)\s]^{(N'-1)/4N'}}}\;. \e
   In the present case, and when $p'_k=p_k$ for all $k$, observe that a simple change of variables does
   not eliminate the $q'$ and $q$ variables from the integrand as was the case for example two. However, we do
   expect that in the continuum limit, the prefactor will again cancel with the coefficients $q'$ and $q$
   from the denominator factor for the simple reason that in the continuum limit, all neighboring $q'$ and
   $q$ values will become equal to ensure a continuous label function $q'(x)$ and $q(x)$, and as such they will
   emerge from the denominator and join the prefactor leading, in the case
   of an infinite spatial volume, to a factor of unity, much as was the case for example two. Thus we expect the
   continuum limit for an infinite volume also to be given by the expression
    \b \<p',q'|p,q\>\hskip-1.5em&&=\lim_{a\ra0}\,M^2\s\int \Pi'_k\s d\p_k\,e^{\t-i\Sigma'_k(p'_k-p_k)\p_k\s a}\\
      &&\hskip1em\times{e^{\t -\half\Sigma'_{k,l}(\p_k/|q'_k|)\s A_{k-l}\s (\p_l/|q'_l|)\s a^{2s}-\half\s W((\p/|q'|)\s a^{(s-1)/2})}}\\
      &&\hskip1em\times {e^{\t -\half\Sigma'_{k,l}(\p_k/|q_k|)\s A_{k-l}\s (\p_l/|q_l|)\s a^{2s}-\half\s W((\p/|q|)\s a^{(s-1)/2})}}\\
      &&\hskip1em\times\frac{1}{[\Pi'_k[\s\Sigma'_l\s J_{k,l}\s\p^2_l)\s]^{(N'-1)/2N'}}\;. \e
      On the other hand, in a finite spatial volume, the preceding expression for the coherent state overlap function would be multiplied by the factor
        \b e^{\t -(1/2V')\tint[\s\ln|q'(x)|+\ln|q(x)|\s]\,d^s\!x}\;,\e
      where $V'\equiv N'\s a^s=(L\s a)^s<\infty$ is the volume of the spatial slice.

      This concludes our discussion of the affine coherent states relevant for
      the field theory models of interest.

\section{Brief Review of the Author's Approach to\\Nonrenormalizable Models}
We now outline the author's program to study the Euclidean-space formal functional integral given (for $\hbar=1$ and $n\ge5$) by
 \b S(h)={\cal N}\int \exp(\s \tint\{h\p-\half[(\nabla\p)^2+m_0^2\s\p^2]-g_0\s\p^4\s\}\,d^n\!x)\,
 {\cal D}\p\;. \label{e1}\e
 A study of such an expression via conventional perturbation theory is not acceptable because of the
 unlimited number of new counterterms needed such as $\p^6,\, \p^8,\, \p^{10},$ $\ldots$, etc., as well as
 higher derivative terms as well. We need a radically new approach.
\subsubsection*{A lattice framework to build on}
 We now sketch our alternative approach. First, we adopt a finite, periodic, hypercubic, spacetime
 lattice with $L<\infty$ sites on a side, a lattice spacing of $a>0$, and lattice sites labeled by a
 multi-integer $k=\{k_0,k_1,k_2,\ldots,k_s\}$, $k_j\in\mathbb{Z}$, where $s=n-1$ is the number of spatial dimensions and $k_0$
 denotes the Euclidean time direction, which will become the true time direction after a Wick rotation. Second,
 we approximate the former equation by a lattice functional integral given by
  \b N\!\int\exp(\s\Sigma_k\{h_k\p_k\s a^n-\half(\p_{k^*}-\p_k)^2\s a^{n-2}-\half\s m_0^2\s\p^2_k\s a^n-g_0\s \p^4_0\s a^n\s\}\s)\,\Pi_k\s d\p_k\;, \e
 where $k^*$ denotes each of the $n$ next nearest neighbors in a positive sense and a summation over such points
 is implicit. As it stands, this lattice expression represents a lattice cutoff of the formal continuum
 functional integral. We need to introduce a counterterm into this expression to account for the needed
 renormalizations that will appear in the continuum limit. Choosing counterterms on the basis of perturbation
 theory is inappropriate, and we need an alternative principle to choose the counterterms.
\subsubsection*{Hard core interactions}
 There are strong reasons to believe that a perturbation theory {\it about the free theory} does not hold true for nonrenormalizable models. In fact, the author has long argued \cite{kla17} that the nonlinear interaction term for nonrenormalizable models acts partially like a hard core, which, in a functional integral formulation, acts to project out certain field histories that would otherwise be allowed by the free theory alone. An argument favoring this explanation is provided by the Sobolev-type inequality
   \b \{\tint \p(x)^4\,d^n\!x\}^{1/2}\le C\{\tint[\s(\nabla\p(x))^2+m^2\s\p(x)^2]\,d^n\!x\}\;, \e
   which for $n\le4$ holds for $C=4/3$, while for $n\ge5$ holds only for $C=\infty$ \cite{book}. In the latter case, this
   means that there are functions $\p(x)$ such that the left hand side diverges while the right hand side is finite. One example of such a function is given by
      \b \p_{singular}(x)=|x|^{-p}\,e^{\t-x^2}\;,\hskip1.7cm n/4\le p < n/2-1\;. \e

 The issues discussed above are even more self evident for a one dimensional system with the classical action
    \b  I=\tint[\s\half({\dot{x}}(t)^2-x(t)^2)-g\s x(t)^{-4}]\,dt\;, \e
    for which, when $g>0$, the paths are unable to penetrate the barrier at $x=0$, and thus the family of theories for $g>0$ also exhibit a hard-core behavior and they do {\it not} converge to the free theory as $g\ra0$.  This hard-core behavior holds in both the classical and quantum theories for this example.

\subsubsection*{Pseudofree models}
   If the interacting theories do not pass to the free theory as the nonlinear coupling constant
   reduces to zero, to what limit do they converge? We have introduced the term {\it pseudofree model}
    to label the limit of the interacting theories when the coupling constant goes to zero. It is the pseudofree theory about which a perturbation exists, if one exists at all, and not about the usual free theory. Our initial goal
   in understanding nonrenormalizable theories is to get a handle on the pseudofree theory, a theory that
   is fundamentally different from the usual free theory.

 As a result of the lack of any connection of the interacting theories with the usual free theory, our procedure to choose the proper counterterm for nonrenormalizable  models will turn out to be
 somewhat circuitous.
\subsubsection*{Role of sharp time averages}
 Let us consider the average of powers of the expression
    \b \Sigma_{k_0}\s F(\p,a)\,a \;,\e
    where $F(\p,a)$ is a function of lattice points all at a fixed value of $k_0$,
 in any lattice spacetime, based on the distribution generated by the exponential of the action, which
 we denote by $\<\s(\,\cdot\,)\s\>$.  We write the
 average of the $p^{\rm th}$ power of this expression as
   \b \<\s[\Sigma_{k_0}\,F(\p,a)\,a]^p\s\>=\Sigma_{k_{0_1},k_{0_2},\ldots,k_{0_p}}\,a^p\,\<\s F(\p_1,a)\,
   F(\p_2,a)\,\cdots\,F(\p_p,a)\s\>\;,\e
   where $\p_j$ here refers to the fact that ``$k_0=j$" in this term. A straightforward inequality leads to
   \b &&|\,\<\s F(\p_1,a)\,F(\p_2,a)\,\cdots\,F(\p_p,a)\s\>\,|\no\\
   &&\hskip1cm \le |\,\<\s [F(\p_1,a)]^p\s\>\,\<\s [F(\p_2,a)]^p\s\>\,\cdots\,\<\s [F(\p_p,a)]^p\s\>\,|^{1/p}\;, \e
   which casts the problem into one at a sharp time. For sufficiently large $L$, it follows that this sharp
   time expression may be given by
      \b \<\s [F(\p,a)]^p\s\>=\int [F(\p,a)]^p\,\Psi_0(\p)^2\,\Pi'_k\s d\p_k\;, \e
      where the integral is taken over fields at a fixed value of $k_0$, $\Psi_0(\p)^2$ denotes the ground state distribution, and $\Pi'_k$ denotes a product over the spatial lattice at a fixed value of
      $k_0$. Thus we have arrived at the
      important conclusion that if the sharp time average is finite, then the full spacetime average is also finite.

      Both the inequality
      noted above and the argument involving the ground state distribution may be found in
      \cite{kla1,kla2,kla3}.
\subsubsection*{Choosing the ground state}
      Attention now turns to finding the ground state, or more to the point, {\it choosing} the ground
      state so that expressions we desire to be finite actually become finite. In particular, this remark means that the ground state is {\it tailored}
      or {\it designed} so
      that those quantities that are divergent in the usual perturbation theory are in fact rendered finite.
      Once the ground state is chosen, one defines the associated lattice Hamiltonian for the system by means of
      the expression
         \b  \H\equiv-\half\s a^{-s}\s\hbar^2\Sigma'_k\,\d^2/\d\p_k^2+\half\s a^{-s}\s\hbar^2\,[1/\Psi_0(\p)]\,\Sigma'_k\,
         \d^2\s\Psi_0(\p)/\d\p_k^2\;, \e
         where $\Sigma'_k$ denotes a sum over a spatial lattice at a fixed value of $k_0$. Note the appearance
         of $\hbar^2$ in both factors. In turn, the lattice Hamiltonian readily leads to the lattice
         action. In summary, we focus first on the desired modification of the ground state for the system,
         which leads to the lattice Hamiltonian, and finally to the desired lattice action.

         It has been argued that the ground state for a free system ($g_0=0$) and with no counterterm
         is clearly a Gaussian. Moreover, such a function leads to divergences for several quantities of
         interest, such as those expressions basic to a mass perturbation, namely, when $F(\p,a)=\Sigma'_k\s\p^2_k\s a^s$.
         It has also been argued that the {\it source} of those divergences can be traced to a
         specific, single factor when the integrals involved are reexpressed in hyper-spherical
         coordinates, which are defined by
          \b &&\p_k\equiv\kappa\s\eta_k\;,\hskip.5cm \Sigma'_k\p^2_k=\kappa^2\;,\hskip.5cm \Sigma'_k\eta^2_k=1\\
           &&\hskip2.6em \kappa\ge0\;,\hskip.5cm -1\le\eta_k\le1\;. \e

          As an example, consider a typical (Gaussian) integral of interest given by
           \b &&K\int [\Sigma'_k\s\p^2_k\s a^s]^p\;e^{\t-\Sigma'_{k,l}\s\p_k\s A_{k-l}\s \p_l\,a^{2s}}\;\Pi'_kd\p_k\no\\
           &&\hskip.6cm=2K\int\s \k^{2p}\s a^{sp}\;e^{\t-\k^2\Sigma'_{k,l}\eta_k\s A_{k-l}\s\eta_l\,s^{2s}}
           \;\k^{(N'-1)}\,d\k\,\delta(1-\Sigma'_k\eta^2_k)\,\Pi'_kd\eta_k\;.   \e
           Here, $A_{k-l}\propto a^{-(1+s)}$ is a matrix responsible for the spatial gradient and other suitable quadratic terms in the lattice Hamiltonian, and $N'$ is the total number of lattice sites in a
           spatial slice of the lattice. In the continuum limit, in which $a\ra0$, it follows that
           $L\ra\infty$ in such a way, initially, that $V'\equiv(L\s a)^s=N'\s a^s<\infty$. Later, one may
           extend the continuum limit procedure so that $V'\ra\infty$ as well. In short, in the
           continuum limit, it follows that $N'\ra\infty$. It is not difficult to show that in the
           foregoing integral, as displayed in hyper-spherical coordinates, it is the term
           $N'$ in the measure factor $\k^{(N'-1)}$ that is {\it the very source of the divergences!}
           If one could change that factor to one that remains {\it finite} in the continuum limit,
           {\it the divergences would disappear!}

           How is one to change a factor that has arisen from a bona-fide coordinate transformation?
           The answer is: It cannot be done {\it directly}, but it can be done {\it indirectly}. The way
           to do so is to choose a different, non-Gaussian ground state distribution, one that has roughly the form
             \b  K'\,\frac{1}{\k^{(N'-1)}}\,e^{\t-\Sigma'_{k,l}\,\p_k\s A_{k-l}\s \p_l\,a^{2s}}\;, \e
             a form that has an additional factor in the denominator to cancel the measure factor $\k^{(N'-1)} $
             altogether; in fact, it doesn't need to cancel it all, but it is a reasonable place to
             begin. Such a ground state distribution arises from a Hamiltonian that is not quadratic
             but contains another component that we identify as the desired counterterm. Finally,
             that counterterm is taken over to the lattice action, and thereby we have determined
             the counterterm in this convoluted manner!

             There are many ways to choose the Hamiltonian so that the counterterm leads to a modification
             of the ground state of the desired form. A large class of ground state modifications may be given by
               \b M\frac{1}{\Pi'_k[\Sigma'_l\s J_{k,l}\s\p^2_l\s]^{(N'-1)/4\s N'}}
               \;e^{\t-\half\Sigma'_{k,l}\,\p_k\s A_{k-l}\s\p_l-\half\s W(\p\s a^{(s-1)/2})} \e
               for various choices of the constant coefficients $J_{k,l}$. For example, one may choose $J_{k,l}=
               \delta_{k,l}$, and this is appropriate to discuss ultralocal models, which may be
               described as relativistic models with their spatial gradient terms removed.
               As we have shown elsewhere, such a choice leads to a Poisson ground state distribution,
               which, although appropriate and correct for ultralocal models, is not desirable for
               truly relativistic models. For relativistic models, on the other hand, it has been
               proposed \cite{klaold} to choose the expression
                 \b J_{k,l}=\frac{1}{2 s+1}\delta_{k,l\s\in\s\{k\s\cup\s {\rm nn}\}}\;, \e
                 where the expression $l\in\{k\s\cup \s{\rm nn}\}$ means that $l=k$ and all the spatial nearest neighbors to $k\s$; $J_{k,l}=0$ elsewhere in the spatial slice, and $\Sigma'_l\s J_{k,l}=1$.

                 With $J_{k,l}$ so chosen, it follows that the Hamiltonian does not represent a local
                 interaction in the continuum limit due to cross terms coming from one derivative each of
                 the $A$ and $J$ terms. To fix that, the unwanted cross terms are removed from the lattice
                 Hamiltonian by means of a suitable, auxiliary term $W(\p\s a^{(s-1)/2})$ in the ground state that
                 effects mid-level field values. Readers who may be interested in what form
                 of counterterm for the lattice action such a modified ground state leads to should consult
                 \cite{kla1,kla2,kla3}.

                 This is not the place to debate the merits of the suggested proposal for the relativistic models
                 and their proposed ground state wave functions. Rather, in this paper, we accept the suggested
                 form of the ground state, and, consequently, we are then led to the
                  set of affine coherent states with the indicated
                 ground state chosen as the fiducial vector that was discussed in the previous section.
\section{Commentary}
       The discussion of affine coherent states for a single degree of freedom emphasized the difficulty in
       ensuring that $\<p,q|\s Q\s|p,q\>=q$ for all $q\in \mathbb{R}\setminus0$. The solution involved a superposition
       of disjoint states. When generalized to infinitely many degrees of freedom, this disjoint feature would have
  led to an infinite number of unitarily inequivalent representations of the affine variables. To avoid this
  situation, we instead accepted the requirement that $\<p,q|\s Q^2\s|p,q\>=q^2$, a compromise which eventually led to a suitable representation for infinitely many degrees of freedom. (It is interesting to observe
  that this modification has some similarities with the Wilson construction for wavelets which also involves
  symmetric fiducial vectors; e.g., see \cite{dau91}.)

  For quantum gravity, which is the principal nonrenormalizable model of interest, it is noteworthy that the
  classical field variable $g_{a\s b}(x)$, the $3\times3$ spatial metric, forms a positive-definite matrix and
  thus it is possible to define affine coherent states for quantum gravity such that
    \b \<\pi,g|\s \hat{g}_{a\s b}(x)\s|\pi,g\>=g_{a\s b}(x) \e
    with an irreducible representation of the appropriate affine variables; see \cite{klaAA}. This fact
    means that for the gravitational field we have the best situation we could hope for!


\begin{thebibliography}{99}
\bibitem{kla17} J.R. Klauder, ``Remarks on Nonrenormalizable Interactions",
Phys. Lett. B {\bf 47}, 523-525 (1973).

\bibitem{aron}Aronszajn, N.: ``La th\'eorie des noyaux reproduisants et ses applications, Première Partie'', Proc. Cambridge Phil. Soc. {\bf 39}, 133--153 (1943); ``Note additionnelle à l'article 'La th\'eorie des noyaux reproduisants et ses applications'', Proc. Cambridge Phil. Soc. {\bf 39}, 205--205 (1943);
    ``Theory of Reproducing Kernels'', Transactions of the American Mathematical Society {\bf 68},
    337-404 (1950).

\bibitem{book} J.R. Klauder, {\it Beyond Conventional Quantization}, (Cambridge
University Press, Cambridge, 2000 \& 2005).

\bibitem{kla1} J.R. Klauder, ``Divergence-free Nonrenormalizable
Models'', { J. Phys. A: Math. Theor.} {\bf 41}, 335206 (15pp) (2008).

\bibitem{kla2} J.R. Klauder, ``Taming Nonrenormalizability'', J. Phys. A: Math. Theor. {\bf 42},  335208 (7pp)
(2009).

\bibitem{kla3} J.R. Klauder, ``Rethinking Renormalization'', arXiv:0904.2869.

\bibitem{klaold} J.R. Klauder, ``A New Approach to Nonrenormalizable Models'',
{Ann. Phys.} {\bf 322}, 2569-2602 (2007).

\bibitem{dau91} I. Daubechies, S. Jaffard, and J.L. Journ\'e, ``A Simple Wilson Orthonormal Basis with
Exponential Decay'', SIAM J. Math. Anal. {\bf 22}, 554-572 (1991).

\bibitem{klaAA} J.R. Klauder, ``Noncanonical Quantization of Gravity. I. Foundations of Affine Quantum Gravity'', J. Math. Phys. {\bf 40}, 5860-5882 (1999); ``Noncanonical Quantization of Gravity. II. Constraints and the Physical Hilbert Space'', J. Math. Phys. {\bf 42}, 4440-4464 (2001).


\end{thebibliography}
\end{document}